\pdfoutput=1
\newcommand*{\ATLASLATEXPATH}{latex/}
\documentclass[cernpreprint,texlive=2011,txfonts,UKenglish]{\ATLASLATEXPATH atlasdoc}

\usepackage{\ATLASLATEXPATH atlaspackage}
\usepackage{multirow}
\usepackage{\ATLASLATEXPATH atlasbiblatex}
\usepackage{\ATLASLATEXPATH atlascontribute}
\usepackage{\ATLASLATEXPATH atlasphysics}

\graphicspath{{logos/}{figures/}}

\AtlasTitle{
Search for the Standard Model Higgs boson produced by vector-boson fusion and decaying to bottom quarks 
in $\sqrt{s} = 8$~$\tev$ $pp$ collisions with the ATLAS detector
}

\author{The ATLAS Collaboration}
\AtlasRefCode{HIGG-2014-12}
 \PreprintIdNumber{CERN-EP-2016-076}

\AtlasJournal{JHEP}

\AtlasJournalRef{JHEP 11 (2016) 112}
\AtlasDOI{10.1007/JHEP11(2016)112}

\AtlasAbstract{
A search with the ATLAS detector is presented for the Standard Model Higgs boson 
produced by vector-boson fusion and decaying to a pair of bottom quarks, 
using 
20.2~${\rm fb}^{-1}$ of LHC proton--proton collision data at ${\sqrt{s}}$ = 8~$\TeV$.
The signal is searched for as a resonance in the invariant mass distribution of 
a pair of jets containing  $b$-hadrons in vector-boson-fusion candidate events.
The yield is measured to be $-0.8 \pm 2.3$ times the Standard Model cross-section for   
a Higgs boson mass of 125~\GeV. The upper limit on the cross-section
times the branching ratio is found to be 
4.4 times the Standard Model cross-section at the 95\% confidence level, consistent with the expected limit value 
of 5.4 (5.7) in the background-only (Standard Model production) hypothesis.
}

\hypersetup{pdftitle={ATLAS document},pdfauthor={The ATLAS Collaboration}}

\begin{document}

\maketitle

\tableofcontents

\section{Introduction}\label{sec_introduction}

Since the ATLAS and CMS collaborations 
reported 
the observation~\cite{Aad:2012tfa, Chatrchyan:2012ufa} 
of a new particle with a mass of about 125~$\gev$ and with properties consistent with those expected 
for the Higgs boson in the Standard Model (SM)~\cite{Englert:1964et, Higgs:1964pj, Higgs:1964ia}, 
more precise measurements have strengthened the hypothesis that the new particle is indeed 
the Higgs boson~\cite{Aad:2013wqa, Aad:2013wqa2, Aad:2013xqa, Khachatryan:2014jba, Khachatryan:2014kca}. 
These measurements were performed primarily in the bosonic decay modes of the 
new particle: $H \rightarrow \gamma\gamma$, $ZZ$, $W^+W^-$. It is essential to study whether it 
also directly decays into fermions as predicted by the SM. 
Recently CMS and ATLAS reported evidence for the $H \rightarrow \tau^+\tau^-$ decay mode 
at a significance level of 3.4 and 4.5 standard deviations, 
respectively~\cite{CMS-Higgs-taus, CMS-Higgs-fermions, Aad:2015vsa}, and 
the combination of these results qualifies as an observation~\cite{Higgs-taus}.
However, the $H \rightarrow b\bar b$ decay mode has not yet been 
observed~\cite{Tevatron-Higgs-bb, Aad:2014xzb, Chatrchyan:2013zna, Khachatryan:2015bnx, CMS-Higgs-ttbar, ATLAS-Higgs-ttbar},
and the only direct evidence of its existence so far has been obtained by the CDF and D0 collaborations~\cite{Tevatron-Higgs-bb} 
at the Tevatron collider.

The production processes of Higgs bosons at the LHC include gluon fusion ($gg \rightarrow H$, 
denoted ggF), vector-boson fusion ($qq \rightarrow qqH$, denoted VBF), Higgs-strahlung 
($q\bar q' \rightarrow WH, ZH$, denoted $WH/ZH$ or jointly $VH$), and 
production in association with a top-quark pair 
($gg \rightarrow t\bar t H$, denoted $t\bar t H$). While an inclusive 
observation of the SM Higgs boson decaying to a $b\bar b$ pair is difficult in hadron 
collisions because of the overwhelming background from multijet production, 
the $VH$, VBF, and $t\bar t H$ processes offer viable options for the observation of 
the $b\bar b$ decay channel. As reported in 
Refs.~\cite{Aad:2014xzb, Chatrchyan:2013zna, Khachatryan:2015bnx, CMS-Higgs-ttbar, ATLAS-Higgs-ttbar}, 
the leptonic decays of vector bosons, the kinematic properties of the production process, 
and the identification of top quarks
are used to reduce the background for $VH$, VBF, and $t\bar t H$, respectively.

\begin{figure}[!h]
  \centering
  \includegraphics[width=0.45\textwidth]{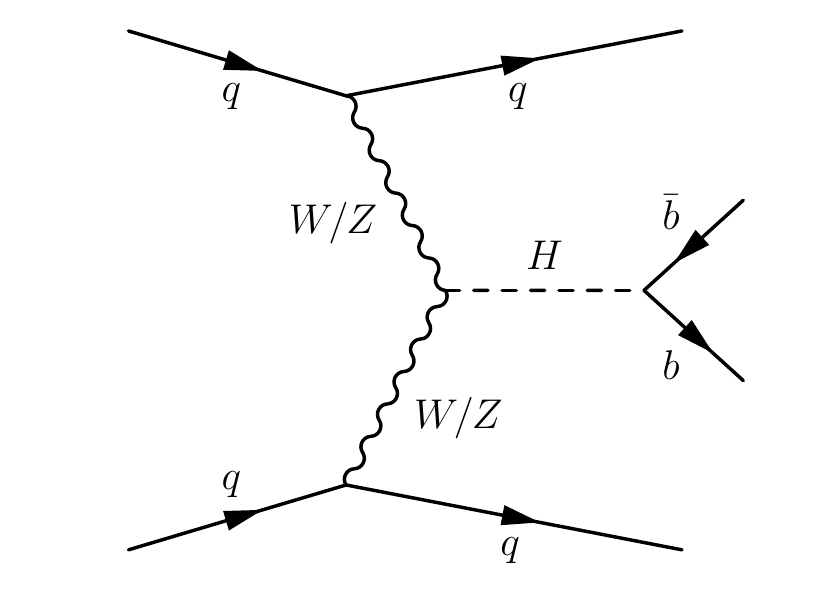}
  \caption{
    An example Feynman diagram illustrating vector-boson-fusion production of the Higgs 
    boson and its decay to a $\bbbar $ pair.
  }
  \label{fig:VBFHiggs-Diagram}
\end{figure}

This article presents a search for VBF production of the SM Higgs boson in the 
$b\bar b$ decay mode (VBF signal or VBF Higgs hereafter)
using data recorded with the ATLAS detector in proton--proton 
collisions at a centre-of-mass energy $\sqrt{s} = 8$~$\TeV$.
The signal is searched for as a resonance in the invariant mass distribution ($m_{bb}$)
of a pair of jets containing $b$-hadrons ($b$-jets) in vector-boson-fusion candidates.
Events are selected by requiring 
four energetic 
jets generated from the $qqH \rightarrow qqb\bar b$ process as illustrated in 
Figure~\ref{fig:VBFHiggs-Diagram}: two light-quark jets (VBF jets) at a small
angle with respect to the beam line
and two $b$-jets from the Higgs boson decay in more central regions. 
Higgs bosons are colour singlets with no colour line 
to the bottom quarks; thus little QCD radiation and hadronic activity is 
expected between the two VBF jets, creating a rapidity gap between them. This feature 
is used to distinguish signal events from multijet events, which form the 
dominant background with a non-resonant contribution to the $m_{bb}$ distribution. 
Another relevant background source arises from the decay of 
a $Z$ boson to $b \bar b$ in association with two jets ($Z \rightarrow b\bar b$ or $Z$ hereafter). 
This results in a resonant contribution to the $m_{bb}$ distribution.

To improve the sensitivity, a multivariate analysis (MVA) is used 
to exploit the topology of the VBF Higgs final state. 
An alternative analysis is performed using kinematic cuts and the $m_{bb}$ distribution. 
The selected sample contains a minor contribution  
from Higgs boson events produced via the ggF process in association with two jets.
These events exhibit an $m_{bb}$ distribution similar to that of VBF Higgs events,
and are treated as signal in this analysis.
The possible contribution of $VH$ production to the signal was also studied but found to be
negligible compared to VBF and ggF Higgs production for this analysis.

\section{The ATLAS detector}\label{sec:AtlasDetector}

The ATLAS experiment uses a multi-purpose particle detector~\cite{ATLASjinst} 
with a forward-backward symmetric cylindrical geometry and a near $4\pi$ coverage in 
solid angle.\footnote{ATLAS uses a right-handed coordinate system with its origin 
at the nominal interaction point (IP)
in the centre of the detector and the $z$-axis along the beam pipe. The $x$-axis points from
the IP to the centre of the LHC ring, and the $y$-axis points upwards. Cylindrical coordinates
$(r,\phi)$ are used in the transverse plane, $\phi$ being the azimuthal angle around the $z$-axis.
The pseudorapidity is defined in terms of the polar angle $\theta$ as $\eta = -\ln \tan(\theta/2)$.
Angular distance is measured in units of $\Delta R \equiv \sqrt{(\Delta\eta)^{2} + (\Delta\phi)^{2}}$.} 
It consists of an inner tracking detector 
(ID) surrounded by a thin superconducting solenoid providing a \SI{2}{\tesla} magnetic 
field, electromagnetic and hadronic calorimeters, and a muon spectrometer (MS).
The ID consists of silicon pixel and microstrip tracking detectors covering the pseudorapidity 
range $|\eta| < 2.5$, and a transition radiation detector in the region $|\eta| < 2.0$.
Lead/liquid-argon (LAr) sampling calorimeters in the region $|\eta| < 3.2$ provide 
electromagnetic energy measurements with high granularity. A hadron 
(steel/scintillator-tile) calorimeter covers the range $|\eta| < 1.7$. The end-cap and 
forward regions are instrumented with LAr calorimeters for both the electromagnetic and 
hadronic energy measurements up to $|\eta| = 4.9$. The MS surrounds the calorimeters 
and is based on three large air-core toroid superconducting magnets with eight coils each.
It includes a system of tracking chambers covering $|\eta| < 2.7$ and fast detectors for 
triggering in the range $|\eta| < 2.4$. The ATLAS trigger system~\cite{ATLAS-Trigger} 
consists of three levels: the first (L1) is a hardware-based system, and the second and 
third levels are software-based systems which are collectively referred to as the high-level trigger (HLT).

\section{Data and simulation samples}\label{sec:Samples}

The data used in this analysis were collected by the ATLAS experiment at a centre-of-mass energy 
of 8~$\TeV$ during 2012, and correspond to an integrated luminosity of 20.2~$\ifb$ 
recorded in stable beam conditions and with all relevant sub-detectors providing high-quality data.

Events are primarily selected by a trigger requiring four jets with transverse momentum 
$\pt>15$~$\gev$ at L1 and $\pt>35$~$\gev$ in the HLT, two of which  
must be identified as $b$-jets by a dedicated HLT $b$-tagging algorithm (HLT $b$-jets).  
This trigger was available during the entire 2012 data-taking period. 
Two triggers designed to enhance the acceptance for VBF $H \rightarrow b\bar b$ events (VBF Higgs triggers) 
were added 
during the 2012 data-taking period. They require either three L1 jets with $\pt>15$~$\gev$ where 
one jet is in the forward region ($|\eta|>3.2$), or two L1 jets in the forward region with $\pt>15$~$\gev$.
These criteria are completed by the requirement of at least one HLT $b$-jet with $\pt>35$~$\gev$.
The VBF Higgs triggers were used for a data sample corresponding to an integrated luminosity of $4.4$~\ifb, 
resulting in an approximately 25\% increase of the signal acceptance.

VBF and ggF Higgs boson signal events and $Z$ boson background events are modelled 
by Monte Carlo (MC) simulations. The signal samples with a Higgs boson mass of $125$~$\gev$ 
are generated by \textsc{Powheg}~\cite{Nason:2004rx, Frixione:2007vw, Alioli:2010xd}, 
which calculates the VBF and ggF Higgs production processes up to next-to-leading order (NLO) 
in $\alpha_{\rm S}$. Samples of $Z$ boson + jets events are generated using \textsc{MadGraph}5~\cite{Alwall:2014hca},
where the associated jets are produced via strong or electroweak (EW) processes including VBF,
and the matrix elements are calculated for up to and including three partons at leading order.
For all simulated 
samples, the NLO CT10 parton distribution functions (PDF)~\cite{Lai:2010vv} are used.
The parton shower and the hadronisation are modelled by \textsc{Pythia}8~\cite{Sjostrand:2007gs},
with the AU2 set of tuned parameters~\cite{ATLAS:2011zja, ATLAS:2011gmi} for the underlying event.

The VBF Higgs predictions are normalised to a cross-section calculation that includes 
full NLO QCD and EW corrections 
and approximate next-to-next-to-leading-order (NNLO) QCD corrections~\cite{Handbook3}.
The NLO EW corrections also affect the $\pt$ shape of the Higgs boson~\cite{Handbook2}.
The $\pt$ shape is reweighted, based on the shape difference 
between \textsc{Hawk}
calculations without and with NLO EW corrections included~\cite{Ciccolini-1, Ciccolini-2}.

The overall normalisation of the ggF process is taken from a calculation at NNLO in QCD 
that includes soft-gluon resummation up to next-to-next-to-leading logarithmic terms 
(NNLL)~\cite{Handbook3}. Corrections to the shape of the generated $\pT$ distribution of 
Higgs bosons are applied to match the distribution from the NNLO calculation with the NNLL 
corrections provided by the \textsc{Hres} program~\cite{Florian-Ferrera, Grazzini-Sargsyan}. 
In this calculation, the effects of finite masses of the top and bottom quarks are included 
and dynamic renormalisation and factorisation scales are used. A reweighting is derived such that 
the inclusive Higgs $\pT$ spectrum matches the \textsc{Hres} prediction, 
and the Higgs $\pT$ spectrum of events with at 
least two jets matches the the \textsc{Minlo} \textsc{hjj}~\cite{MINLO-HJJ} 
prediction, the most recent calculation in this phase space.

The ATLAS simulation~\cite{Aad:2010ah} of the detector is used for all MC events based
on the 
\textsc{Geant}4 program~\cite{Agostinelli:2002hh} except for the response of the calorimeters, 
for which a parameterised simulation~\cite{atlas-fastsim} is used. All simulated events are 
generated with a range of minimum-bias interactions overlaid on the hard-scattering 
interaction to account for multiple $pp$ interactions that occur in the same or neighbouring 
bunch crossings (pile-up). 
The simulated events are processed with the same reconstruction 
algorithms as the data. Corrections are applied to the simulated samples to account for 
differences between data and simulation in the trigger and reconstruction efficiencies 
and in pile-up contributions.

\section{Object reconstruction}\label{sec:ObjectReconstruction}

Charged-particle tracks are reconstructed with a $\pt$ threshold of 400~\MeV.
Event vertices are formed from these tracks and are required to have at least three tracks.
The primary vertex is chosen as the vertex with the largest $\Sigma ~\pt^2$ of the associated tracks.

Jets are reconstructed from topological clusters of energy deposits, after noise suppression, in the 
calorimeters~\cite{AtlasJets} using the anti-$k_t$ algorithm~\cite{Cacciari:2008gp} 
with a radius parameter $R$ = 0.4. Jet energies are corrected for the contribution of pile-up interactions 
using a jet-area-based technique~\cite{Cacciari:2007fd}, and calibrated using $\pt$- and $\eta$-dependent
correction factors determined from MC simulations and in-situ data measurements of $Z$+jet, $\gamma$+jet and 
multijet events~\cite{Aad:2011he, Aad:2014bia}.
To suppress jets from pile-up interactions, which are mainly at low \pt, 
a jet vertex tagger~\cite{jet-vertex-tagger}, based on tracking and
vertexing information, is applied to jets with $\pT<50$~$\gev$ and $|\eta| < 2.4$.

The $b$-jets are identified ($b$-tagged) by exploiting the relatively long lifetime and 
large mass of $b$-hadrons. The $b$-tagging methods are based on the presence of 
tracks with a large impact parameter 
with respect to the primary vertex, and secondary decay vertices. This information is 
combined into a single neural-network discriminant~\cite{Aad:2015ydr}.
This analysis uses a $b$-tagging criterion that, in simulated $t\bar{t}$ events, 
provides an average efficiency of 70\% 
for $b$-jets and a $c$-jet (light-jet) mis-tag rate less than 20\% (1\%).

\section{Event pre-selection}\label{sec:EventSelection}

Events with exactly four jets, each with $\pt > 50$~$\GeV$ and $|\eta| < 4.5$, are retained. The four 
jets are ordered in $\eta$ such that $\eta_1 < \eta_2 < \eta_3 < \eta_4$. 
The jets associated with $\eta_1$ and $\eta_4$ are labelled as VBF jets (or $J1$ and $J2$). 
The other two jets associated with $\eta_2$ and $\eta_3$ (Higgs jets or $b1$ and $b2$) are 
required to be within the tracker acceptance ($|\eta| < 2.5$), and to be identified as $b$-jets. 
The two Higgs jets must be matched to the HLT $b$-jets for events satisfying the primary trigger; 
for events satisfying the VBF Higgs triggers, one of the two Higgs jets is required to be matched to an HLT $b$-jet.
The $50$~$\GeV$ cut on jet $\pt$ shapes the $m_{bb}$ distribution for non-resonant backgrounds, creating a peak near $130$~$\GeV$, which
makes the extraction of a signal difficult. 
This shaping is removed by requiring the $\pt$ of the $b\bar b$ system to exceed $100$~$\GeV$. 
Table~\ref{tab:sel} summarises 
the acceptances of these pre-selection criteria, for the VBF and ggF Higgs 
MC events~\cite{Handbook3, Djouadi:1997yw} and the $Z$ MC events.

\begin{table} [!h]
  \centering
\caption{Cross-sections times branching ratios (BRs)
  used for the VBF and ggF $H \rightarrow \bbbar$ and $Z \rightarrow \bbbar$ MC generation,
  and acceptances of the pre-selection criteria for simulated samples.
}
\label{tab:sel}

\begin{tabular}{|c||r@{.}l|c|}
\hline
Process & \multicolumn{2}{c|}{Cross-section $\times$ BR [pb]} & Acceptance \\ \hline\hline
VBF $H \rightarrow b\bar b$  &~~~~~~~~~~~~0&9 & 6.9 $\times 10^{-3}$\\ \hline
ggF $H \rightarrow b\bar b$  &~~~~~~~~~~~~11&1 & 4.2 $\times 10^{-4}$\\ \hline
$Z \rightarrow b\bar b$ + 1, 2, or 3 partons  &   ~~~~~~~~~~~~5&9 $\times 10^2$ & 3.1 $\times 10^{-4}$\\
\hline
\end{tabular}
\end{table}

For the pre-selected events, corrections are applied to improve the $b$-jet energy measurements. 
If muons with $\pt > 4$~$\gev$ and $|\eta| < 2.5$ are found within a $b$-jet, the 
four-momentum of the muon closest to the jet axis is added to that of the jet (after correcting 
for the expected energy deposited by the muon in the calorimeter material). Such muons are reconstructed 
by combining measurements from the ID and MS
systems, and are required 
to satisfy tight muon identification quality criteria~\cite{Aad:2014rra}. In addition, a 
$\pt$-dependent correction of up to 5\% is applied to account for biases in the response due to 
resolution effects. This correction is determined from simulated $WH$/$ZH$ events following Ref.~\cite{Aad:2014xzb}.

\section{Multivariate analysis}\label{sec:MVA}

A Boosted Decision 
Tree~\cite{bdt1,bdt2} (BDT) method, as implemented in the Toolkit for Multivariate 
Data Analysis package~\cite{Hocker:2007ht}, is used to exploit the characteristics 
of VBF production. 
The BDT is trained to discriminate between VBF Higgs signal events and non-resonant background 
events modelled using the data in the sideband regions of the $m_{bb}$ distribution
($70<m_{bb}<90$~$\gev$ and $150<m_{bb}<190$~$\gev$).

The input variables of the BDT are chosen to exploit the difference in topologies between 
signal events and background events while keeping them as uncorrelated as possible with $m_{bb}$, 
to ensure that the sideband regions provide a good description of the non-resonant background in the signal region. 
In order of decreasing discrimination power, which is determined by removing variables one by one from the analysis, 
the variables are: 
the jet widths of VBF jets having $|\eta| < 2.1$ 
(the jet width is defined as the \pt-weighted angular distance of the jet constituents from the jet axis, 
and is set to zero if $|\eta| > 2.1$), 
which differs on average for quark and gluon jets;
the scalar sum of the $\pt$ of additional jets with $\pt>20$~$\gev$
in the region $|\eta|<2.5$, $\Sigma p_{\rm T}^{\rm jets}$;
the invariant mass of the two VBF jets, $m_{JJ}$; the $\eta$ 
separation between the two VBF jets, $\Delta\eta_{JJ}$; the maximum $|\eta|$ of the two 
VBF jets, max$(|\eta_{J1}|, |\eta_{J2}|)$; the separation between the $|\eta|$ average 
of the VBF jets and that of the Higgs jets, $(|\eta_{J1}|+|\eta_{J2}|)/2 - (|\eta_{b1}|+|\eta_{b2}|)/2$; 
and the cosine of the polar angle of the cross product of the VBF jets momenta, 
$\cos\theta$, which is sensitive to the production mechanism. 

Figures~\ref{fig:bdtin12_1} and \ref{fig:bdtin12_2} show the distributions 
of the BDT input variables in the data and the simulated samples 
for the VBF $H\rightarrow b\bar b$, ggF $H\rightarrow b\bar b$, and
$Z\rightarrow b\bar b$ events that satisfy the pre-selection criteria.
The BDT responses to the pre-selected data and simulated events are compared
in Figure~\ref{fig:bdtout12}.
As expected, the BDT response to the VBF Higgs signal sample is significantly different from
its response to the data, which are primarily multijet events, and also from its response 
to the $Z$ and ggF Higgs samples.

\begin{figure}[bh!]
 \centering
        \subfloat[]{\includegraphics[width =.5\linewidth]{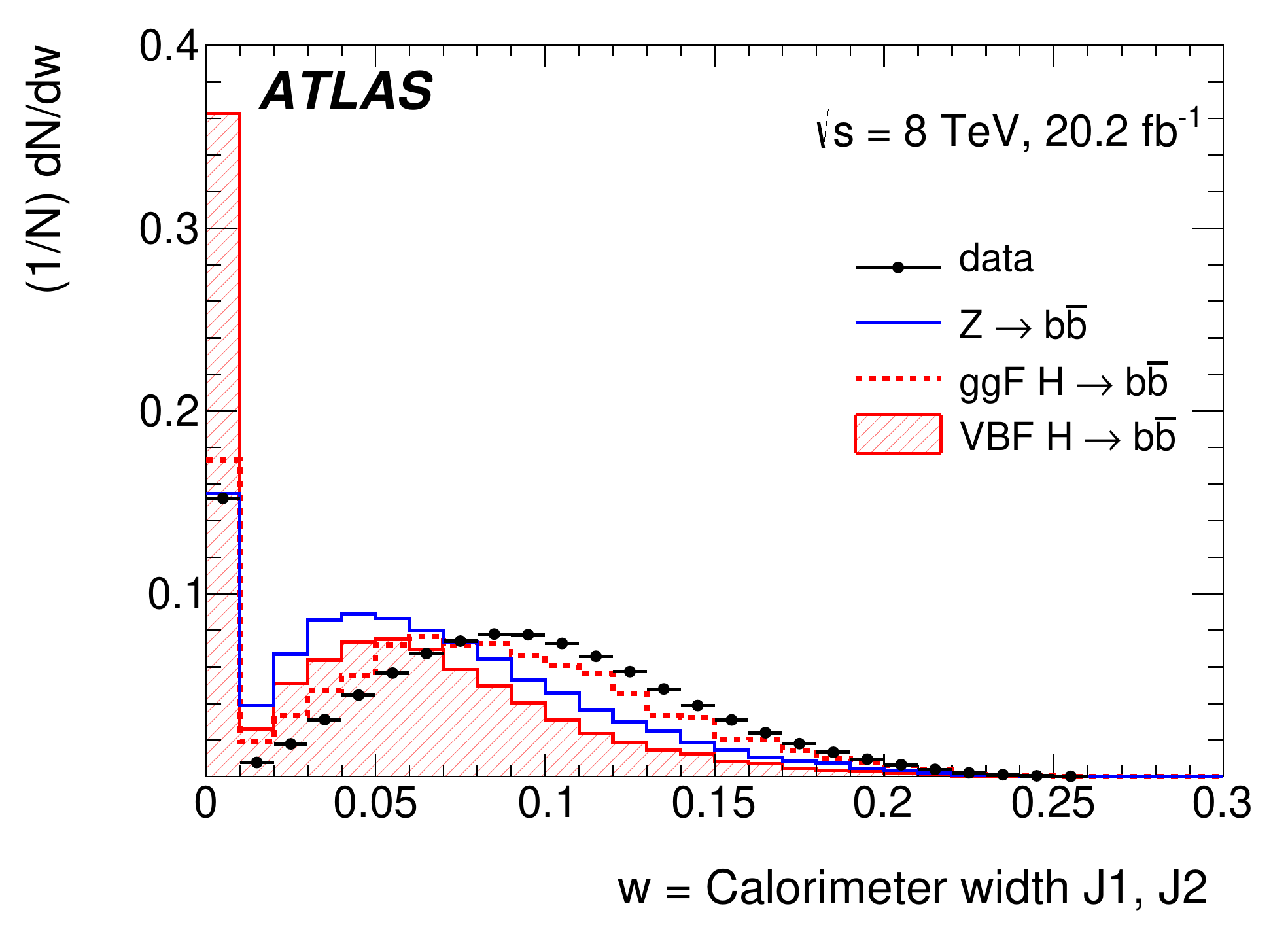}}
\\
                \subfloat[]{\includegraphics[width =.5\linewidth]{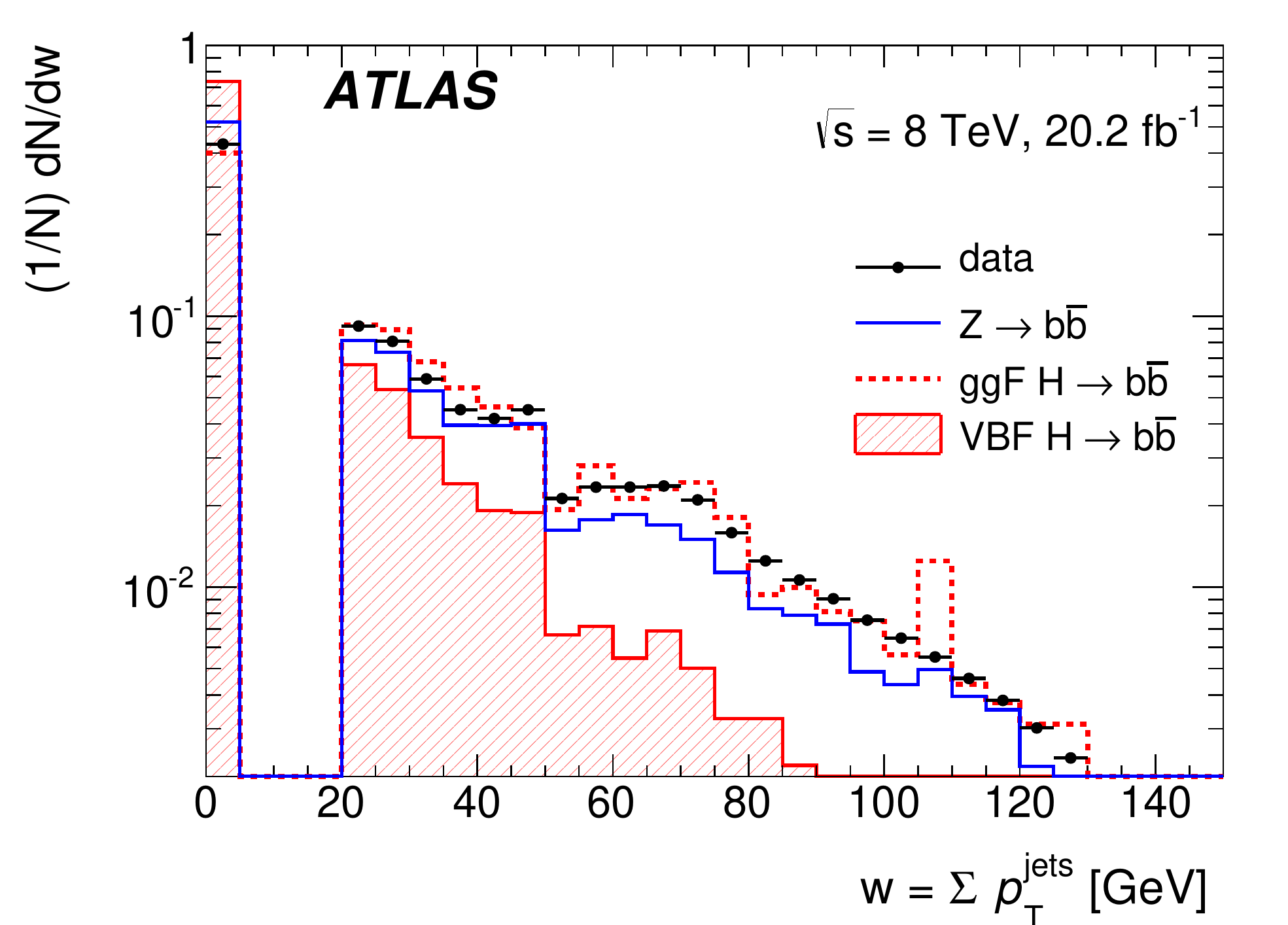}}
        \subfloat[]{\includegraphics[width =.5\linewidth]{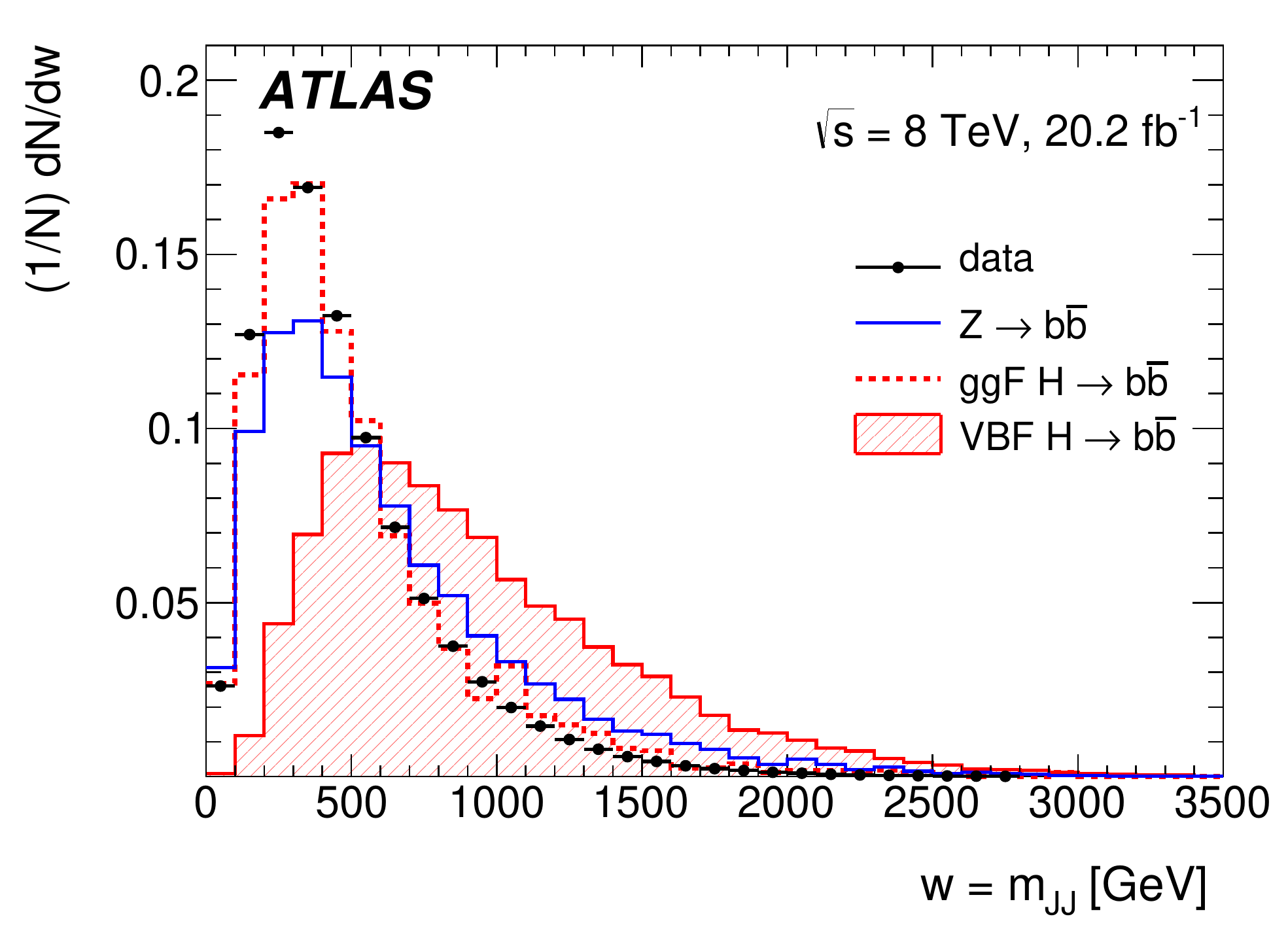}}\\
  \caption{
Distributions of the BDT input variables from the data (points) and the simulated samples
for VBF $H\rightarrow \bbbar$ events (shaded histograms),
ggF $H\rightarrow \bbbar$ events (open dashed histograms)
and $Z\rightarrow \bbbar$ events (open solid histograms). The pre-selection criteria are
applied to these samples. The variables are:
(a) the jet widths  for the VBF jets having $|\eta| < 2.1$ (the jet width is set at zero 
if $|\eta| > 2.1$);
(b) the scalar sum of the $\pt$ of additional jets with $\pt > 20$~$\gev$ 
in the region $|\eta| < 2.5$, $\Sigma p_{\rm T}^{\rm jets}$ 
(the peak at zero represents events without additional jets); and
(c) the invariant mass of the two VBF jets, $m_{JJ}$.
}
  \label{fig:bdtin12_1}
\end{figure}

\begin{figure}[bh!]
 \centering
        \subfloat[]{\includegraphics[width =.5\linewidth]{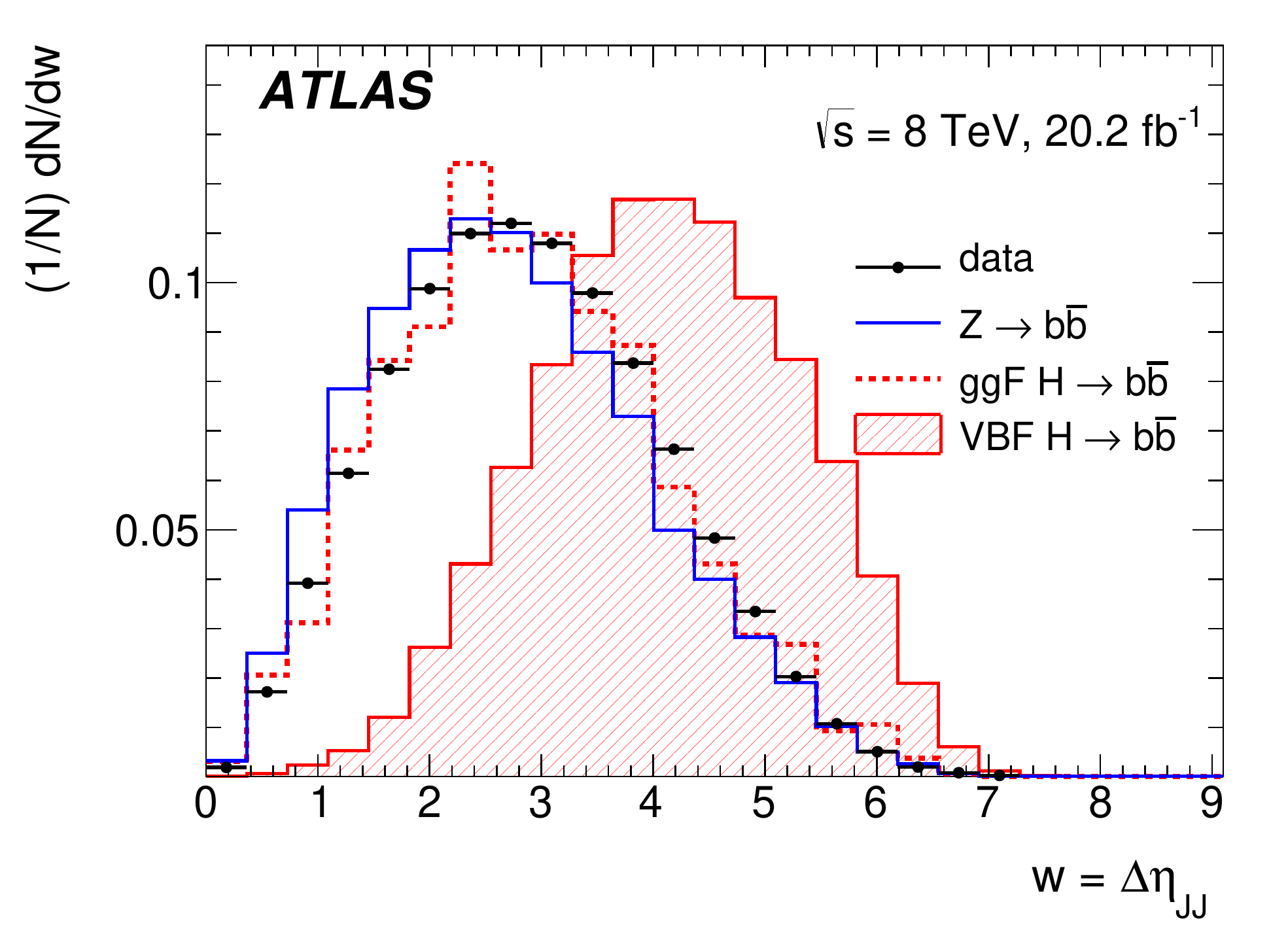}}
        \subfloat[]{\includegraphics[width =.5\linewidth]{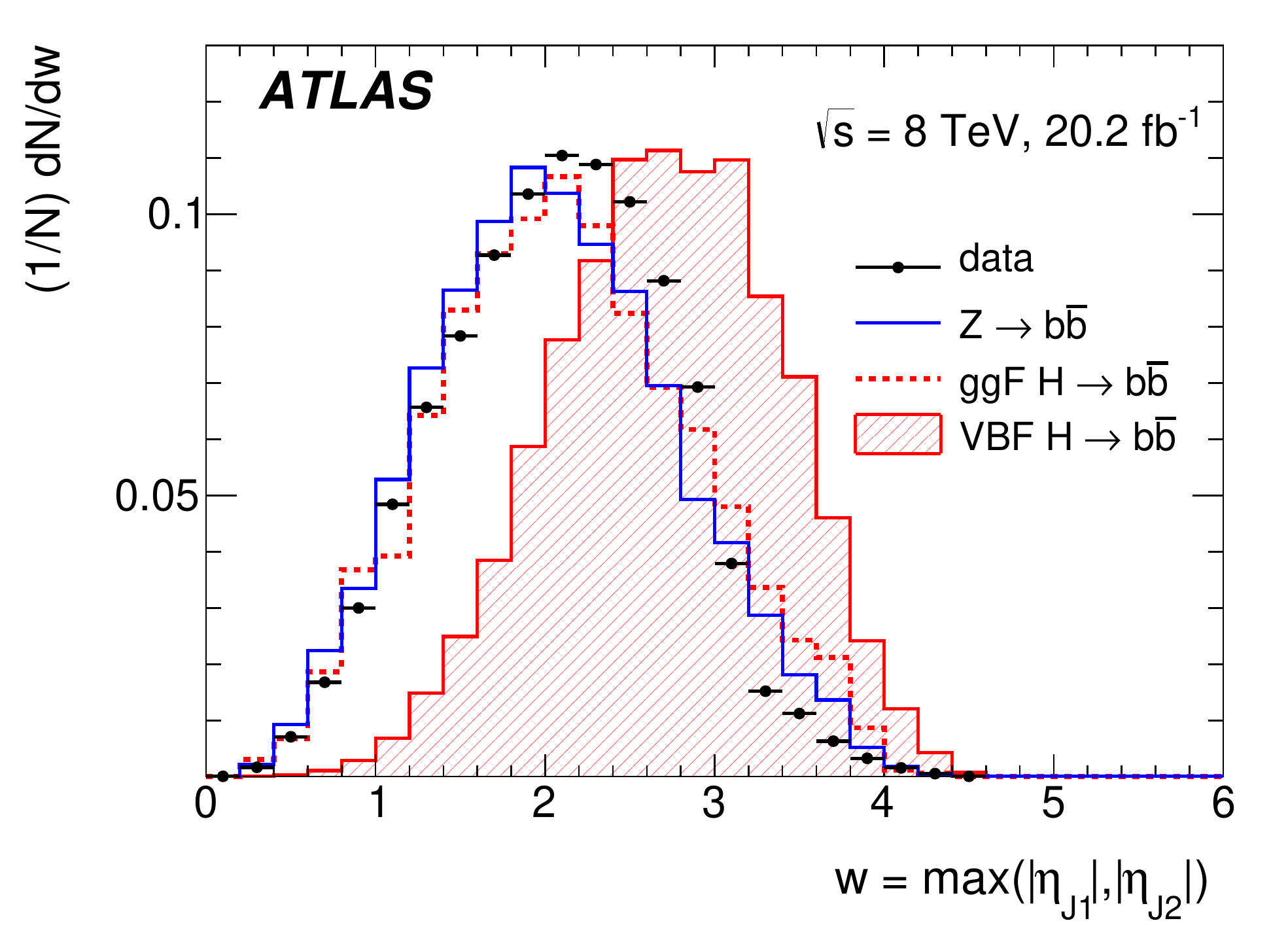}}\\
        \subfloat[]{\includegraphics[width =.5\linewidth]{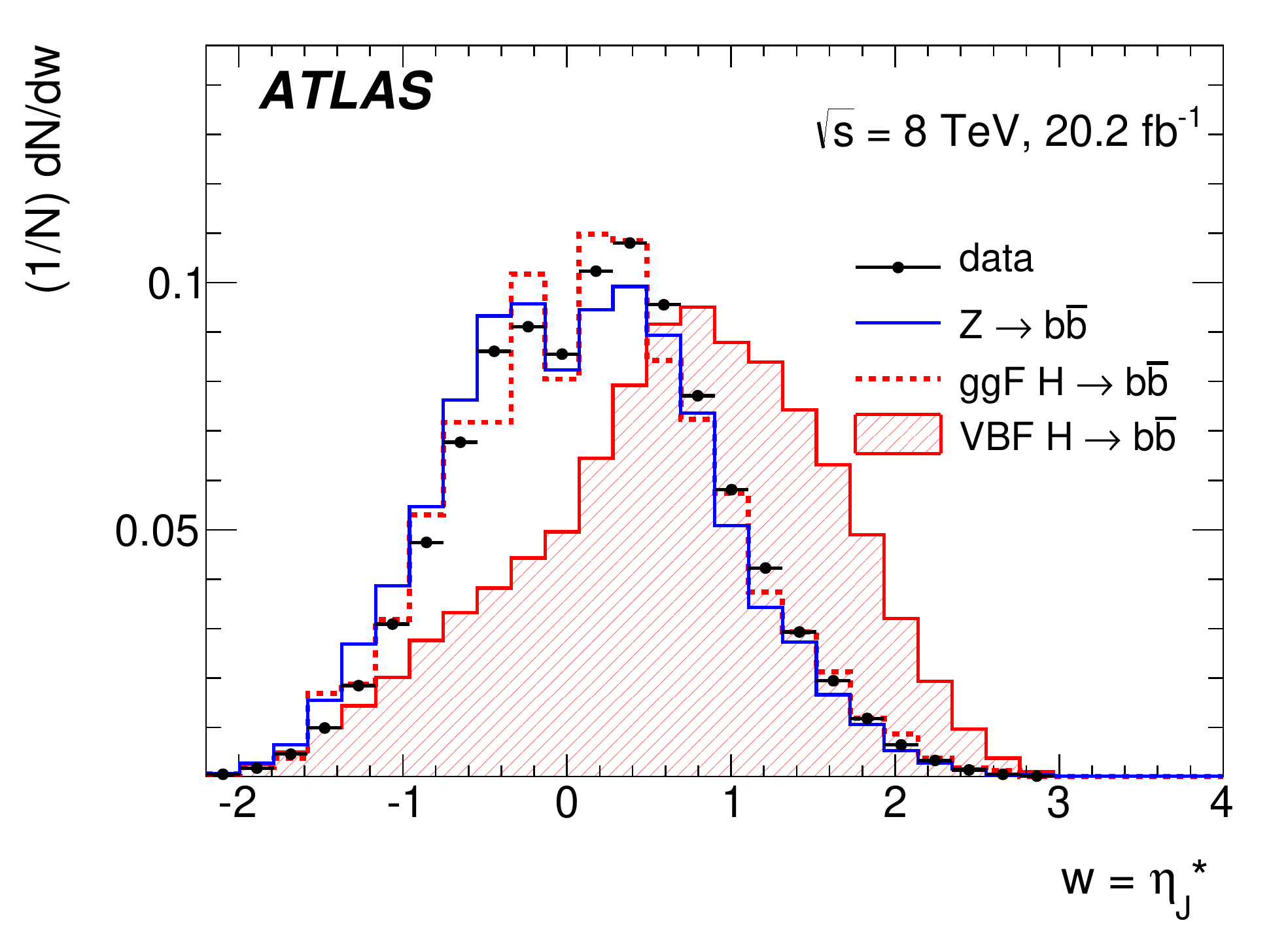}}
        \subfloat[]{\includegraphics[width =.5\linewidth]{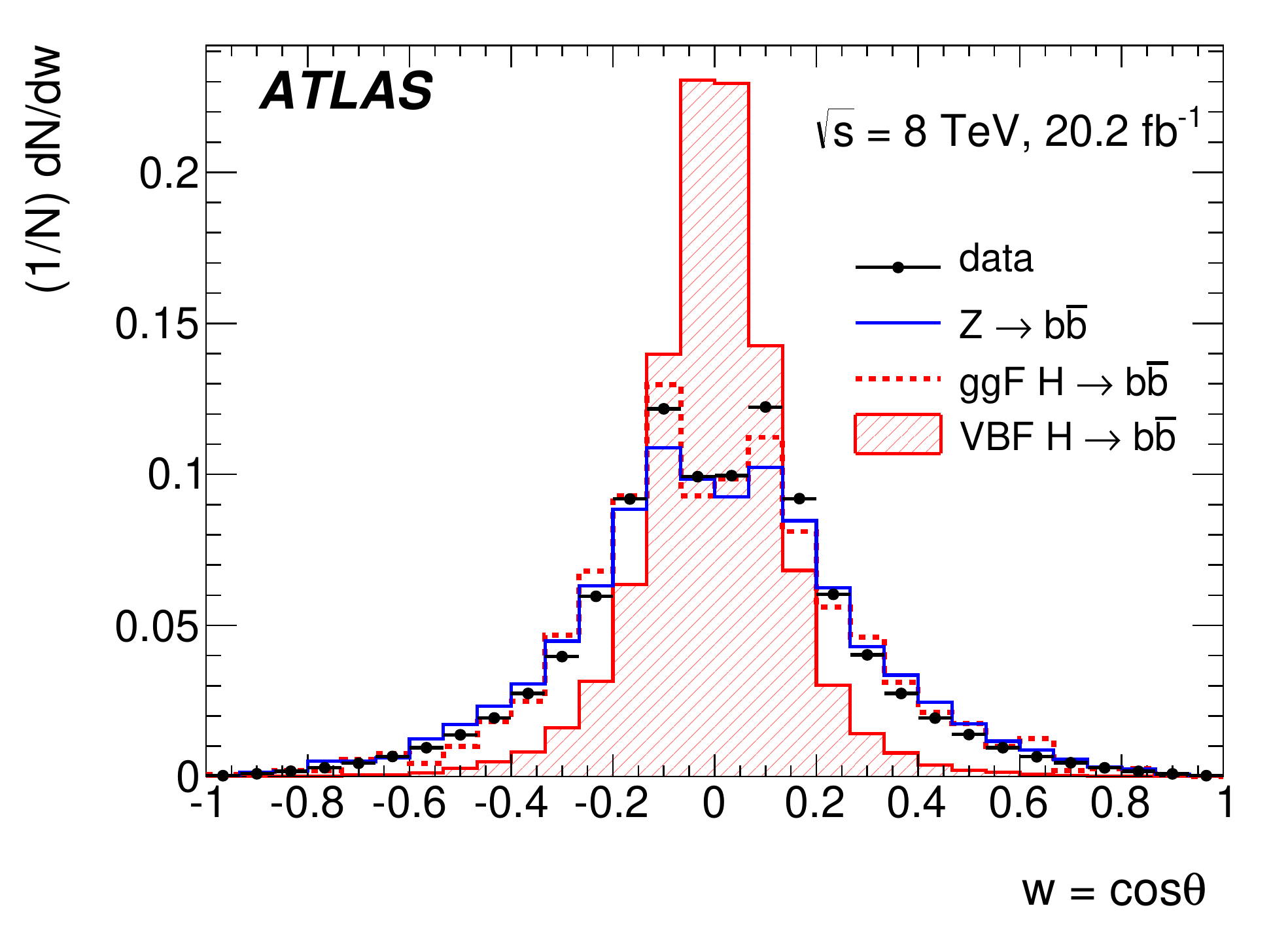}}
  \caption{
Distributions of the BDT input variables from the data (points) and the simulated samples
for VBF $H\rightarrow \bbbar$ events (shaded histograms),
ggF $H\rightarrow \bbbar$ events (open dashed histograms)
and $Z\rightarrow \bbbar$ events (open solid histograms). The pre-selection criteria are
applied to these samples. The variables are:
(a) the $\eta$ separation between the two VBF jets, $\Delta\eta_{JJ}$;
(b) the maximum $|\eta|$ of the two VBF jets, max$(|\eta_{J1}|, |\eta_{J2}|)$;
(c) the separation between the $|\eta|$ average of the VBF jets and that of the Higgs jets,
$\eta_J^* = (|\eta_{J1}|+|\eta_{J2}|)/2 - (|\eta_{b1}|+|\eta_{b2}|)/2$; and
(d) the cosine of the polar angle of the cross product of the VBF jets momenta, $\cos\theta$.
}
  \label{fig:bdtin12_2}
\end{figure}

\begin{figure}[bh!]
 \centering
 \includegraphics[width=0.55\linewidth]{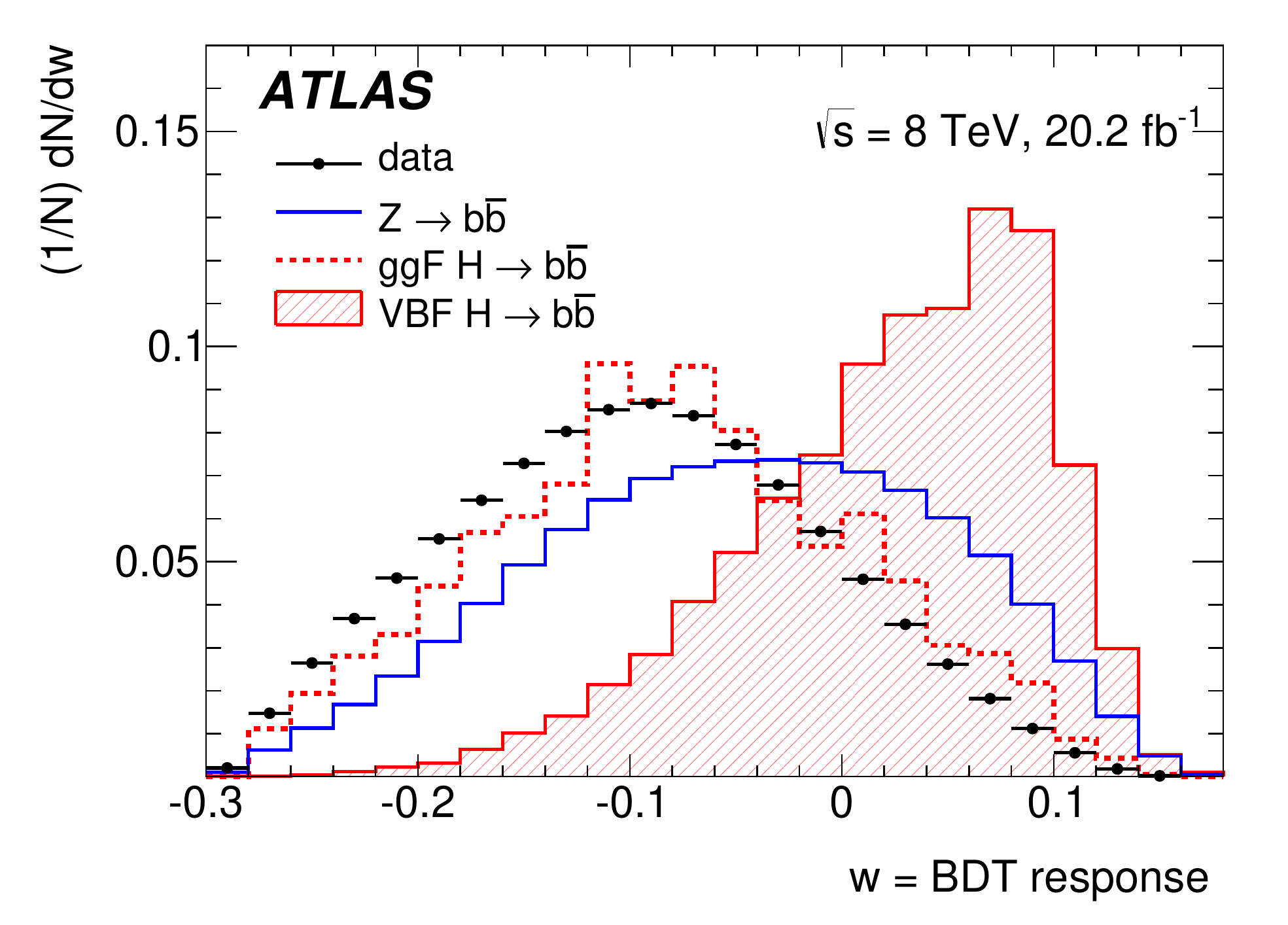}  
  \caption{ 
Distributions of the BDT response to the data (points) and to the simulated samples 
for VBF $H\rightarrow \bbbar$ events (shaded histogram), 
ggF $H\rightarrow \bbbar$ events (open dashed histogram)
and $Z\rightarrow \bbbar$ events (open solid histogram). The pre-selection criteria are 
applied to these samples. 
}
  \label{fig:bdtout12}
\end{figure}

\section{Invariant mass spectrum of the two $b$-jets}\label{sec:Fit}

The signal is estimated using a fit to the $m_{bb}$ distribution 
in the range $70<m_{bb}<300$~$\gev$. The contributions to the distribution 
include $H\rightarrow b \bar b$ events, from either VBF or ggF production; 
$Z\rightarrow b \bar b$ events produced in association with jets; 
and non-resonant processes such as multijet, $t\bar t$, single top, and $W$+jets
production. 
In order to better exploit the MVA discrimination power, the fit is performed 
simultaneously in four categories based on the BDT output.
The boundaries of the four categories, shown in Table~\ref{tab:cat}, 
were optimised by minimising the relative statistical uncertainties, 
$\sqrt{N_{\text{sig}}+N_{\text{bg}}}/N_{\text{sig}}$,
where $N_{\text{sig}}$ and $N_{\text{bg}}$ are the expected numbers of signal and 
background events, respectively.
Table~\ref{tab:cat} shows, for each category, the total number of events observed 
in the data and the number of Higgs events expected from the VBF and ggF production 
processes, along with the number of $Z$ events expected in the entire mass range.
The categories in Table~\ref{tab:cat} are listed in order of increasing sensitivity.

The shapes of the $m_{bb}$ distributions for Higgs and $Z$ boson events are taken 
from simulation.  Their shapes in the four categories are found to be comparable; 
therefore the inclusive shapes are used. The $m_{bb}$ shapes for VBF and ggF Higgs boson events are similar, 
as expected.  
In order to minimise the effects of the limited MC sample size,
the resulting $m_{bb}$ histograms for Higgs and $Z$ events are smoothed 
using the 353QH algorithm~\cite{Hist-smoothing}. 
The $m_{bb}$ distributions used in the fit are shown in Figure~\ref{fig:mbb}.
The Higgs yield is left free to vary. The $Z$ yield is constrained to the SM prediction within
its theoretical uncertainty (see Section~\ref{sec:SystematicsTheory}).

\begin{table}[!h]
\begin{center}
\caption{
 Expected numbers of events for VBF and ggF $H\rightarrow \bbbar$ and 
 $Z\rightarrow \bbbar$ processes, and the observed numbers of events in data
 with $70 < m_{bb} < 300$~$\gev$, after the pre-selection criteria are applied, 
 in the four categories of the BDT response. 
 The categories are listed in order of increasing sensitivity. 
 The values in the parentheses represent the boundaries of each BDT category.
}
\label{tab:cat}
\begin{tabular}{|c||r||r|r|r|r|}
\hline
      \multirow{2}{*}{Process}         
&   \multirow{2}{*}{Pre-selection}         
& Category I      
& Category II    
& Category III       
& Category IV  \\ 
                                                     
&                          
& ($-0.08$ to $0.01$)
& ($0.01$ to $0.06$)
& ($0.06$ to $0.09$)
& ($> 0.09$)
\\ 
               \hline
	       \hline
      VBF $H\rightarrow b\bar b$       &  130 & 39 & 33~~~ & 23~~~ & 19~~~ \\  
      	\hline
      ggF $H\rightarrow b\bar b$       &   94 & 31 & 8.5 & 3.8 & 1.6 \\  
      	\hline
      $Z\rightarrow b\bar b$           & 3700 & 1100 & 350~~~ & 97~~~ & 49~~~ \\  
      	\hline
      Data                             & 554302 & 176073 & 46912~~~ & 15015~~~ & 6493~~~ \\ 
      	\hline
\end{tabular}
\end{center}
\end{table}

\begin{figure}[htb!]
 \centering
 \includegraphics[width=0.55\linewidth]{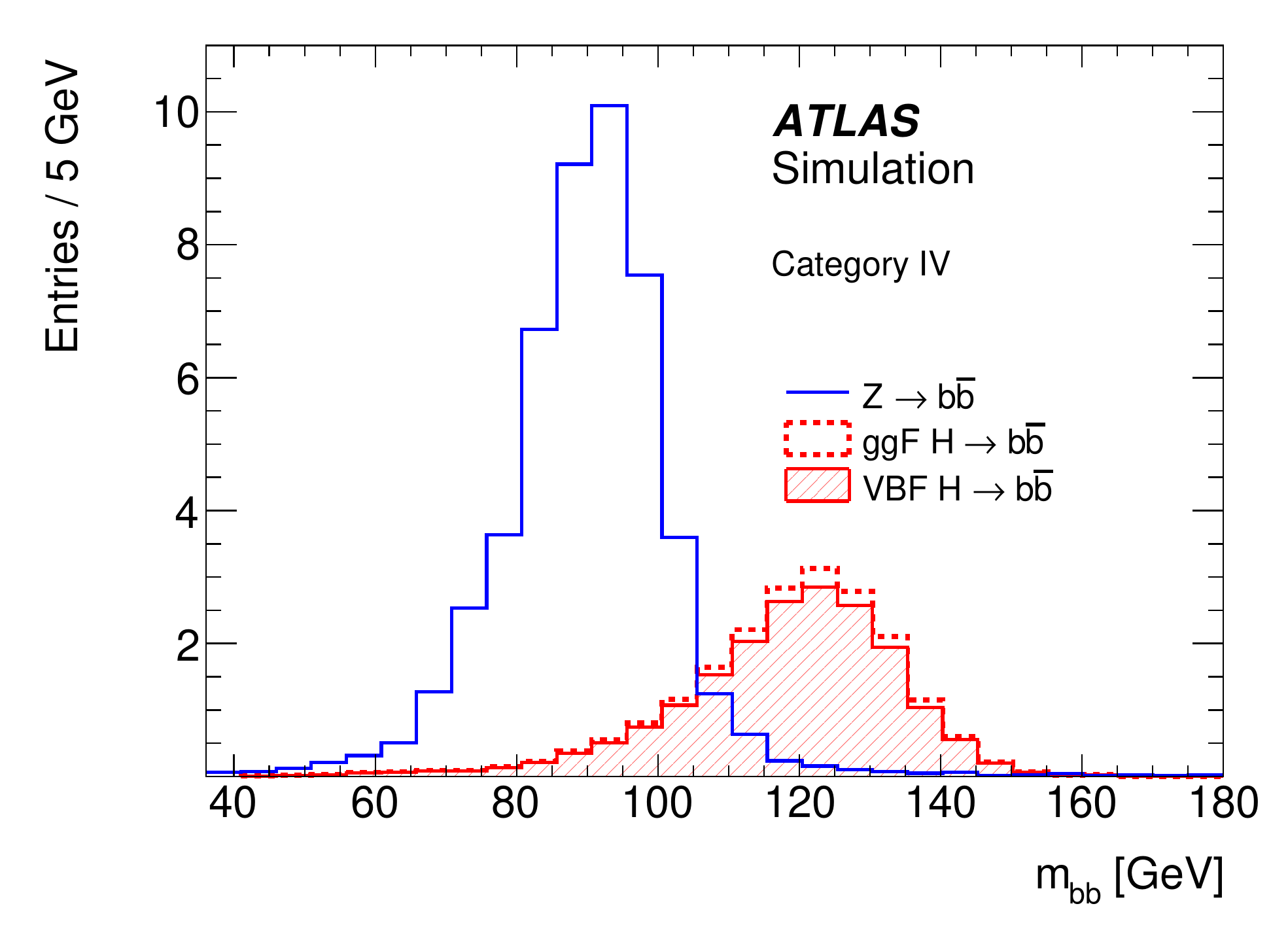}
 \caption{ 
 Simulated invariant mass distributions of two $b$-jets from decays of Higgs bosons, summed for 
 VBF (shaded  histogram) and ggF (open dashed histogram) production, as well as from decays 
 of $Z$ bosons (open solid histogram), normalised to the expected contributions in 
 category IV, which gives the highest sensitivity.
  }
 \label{fig:mbb}
\end{figure}

A data-driven method is used to model the $m_{bb}$ distribution of the non-resonant background. 
Data in the sidebands of the $m_{bb}$ distribution
are fit simultaneously to a function which is then interpolated to the signal region.
The analytic forms considered are Bernstein polynomials~\cite{Bernstein}, combinations 
of exponential functions, and combinations of Bernstein polynomials and exponential 
functions with various numbers of coefficients, and functions with a $\chi^2$ probability 
greater than 0.05, that do not introduce a bias, are selected. For each form, the minimum 
number of coefficients is determined by performing an F-test, 
and the corresponding
function is chosen as a candidate function. The fitted signal 
strength is measured for each candidate function using toy samples.
The function giving the smallest bias is used as the nominal distribution. 
The function giving the second smallest bias is taken as an alternative distribution, 
and is used to estimate the systematic uncertainty due to the choice of analytic function.
The shapes of the $m_{bb}$ distributions are observed to be different in the four categories.
Bernstein polynomials of different degrees, fourth-order in category~I and 
third-order in the higher-sensitivity categories, 
are found to best describe the $m_{bb}$ shape of the non-resonant background. The nominal and 
alternative functions are summarised in Table~\ref{tab:para}.

\begin{table}[!h]
\begin{center}
\caption{
Nominal and alternative functions describing the non-resonant background
in the four BDT categories. The fourth-, third-, and second-order 
Bernstein polynomials are referred to as 4$^{\mathrm{th}}$ Pol., 3$^{\mathrm{rd}}$ Pol., and 
2$^{\mathrm{nd}}$ Pol.
}
\label{tab:para}
\begin{tabular}{|c||c|c|c|c|}
\hline
& category I
& category II
& category III
& category IV  \\
               \hline
               \hline
Nominal 
& 4$^{\mathrm{th}}$ Pol.
& 3$^{\mathrm{rd}}$ Pol.
& 3$^{\mathrm{rd}}$ Pol.
& 3$^{\mathrm{rd}}$ Pol.
\\ \hline
Alternative 
& 2$^{\mathrm{nd}}$ Pol. $\times$ exponential
& 3 exponentials
& 2 exponentials
& exponential
\\ \hline

\end{tabular}
\end{center}
\end{table}

\section{Sources of systematic uncertainty}\label{sec:Systematics}

This section discusses sources of systematic uncertainty: experimental 
uncertainties, uncertainties on the modelling of the non-resonant background, 
and theoretical uncertainties on the Higgs and $Z$ processes. 
The uncertainties can affect the normalisation and the kinematic distributions individually or both together.

\subsection{Experimental uncertainties}\label{sec:SystematicsExp}

The dominant experimental uncertainties on the Higgs signal yield arise from 
the statistical uncertainty due to the finite size of the MC samples, the jet energy scale uncertainty,
and the $b$-jet triggering and tagging, 
contributing 15\%, 10--20\%, and 10\% respectively, to the total uncertainty on the Higgs yield.
Limited MC sizes affect the normalisation via the acceptance of the signal events and 
the shape of the signal $m_{bb}$ distribution.
Several sources contribute to the uncertainty on the jet energy scale~\cite{Aad:2014bia}. 
They include the in situ jet calibration, pile-up-dependent corrections and the flavour 
composition of jets in different event classes. 
The shape of the $m_{bb}$ distribution for the Higgs signal and the $Z$ background is affected 
by the jet energy scale uncertainty. Moreover, the change in the jet energy modifies 
the value of the BDT output and can cause migration of events between BDT categories.
The $b$-jet trigger and tagging efficiencies are another source of systematic uncertainty, 
contributing 10\% to the total uncertainty. 
They are calibrated using multijet events containing a muon and $t\bar t$ events, 
respectively~\cite{btagnote2014b_OF}. 
The uncertainty on the jet energy resolution contributes about 4\%.
The uncertainty on the integrated luminosity, 1.9\%~\cite{Lum-8TeV}, 
is included, but is negligible compared to the other uncertainties mentioned above.

\subsection{Modelling uncertainties on the $m_{bb}$ shape of the non-resonant background}
The uncertainties on the shape of the $m_{bb}$ distribution for the non-resonant background is the largest source 
of systematic uncertainty, contributing about 80\% to the total uncertainty on the Higgs yield. 
The dominant contributions to this source come from the limited number of events in the $m_{bb}$ sidebands of 
the data used for the fit to the nominal function, and from the choice of the function.
For the latter, an alternative function is chosen for each BDT region, as described  
in Section~\ref{sec:Fit} and listed in Table~\ref{tab:para}. 
Pseudo-data are generated using the nominal functions and are fit simultaneously in the four
BDT categories with nominal and alternative functions. 
The bin-by-bin differences in the background yield predicted by the two alternative descriptions
are used to estimate, by means of an eigenvector decomposition, the corresponding systematic
uncertainties.

\subsection{Theoretical uncertainties}\label{sec:SystematicsTheory}
The uncertainties on the MC modelling of the Higgs signal events contribute 
about 10\% to the total uncertainty on the Higgs yield.
The sources for these uncertainties are higher order QCD corrections, the modelling of the underlying event 
and the parton shower, the PDFs, and the $H \rightarrow b\bar b$ branching ratio.
An uncertainty on higher order QCD corrections for the cross-sections and acceptances is estimated 
by varying the factorisation and renormalisation
scales, $\mu_{\rm F}$ and $\mu_{\rm R}$, independently by a factor of two around the nominal values~\cite{Handbook2} 
with the constraint $0.5 \le \mu_{\rm F}/\mu_{\rm R} \le 2$.
Higher order corrections to the $\pt$ spectrum of the Higgs boson (described in Section~\ref{sec:Samples}) are 
an additional source of the modelling uncertainties. This uncertainty is estimated by 
comparing the results between LO and NLO calculations for VBF production and by varying the factorisation 
and renormalisation scales for ggF production.
Uncertainties related to the simulation of the underlying event and the parton shower
are estimated by comparing distributions obtained using 
\textsc{Powheg}+\textsc{Pythia}8 and \textsc{Powheg}+\textsc{Herwig}~\cite{Corcella:2000bw}.
The uncertainties on the acceptance due to uncertainties in the PDFs are estimated by 
studying the change in the acceptance when 
different PDF sets such as MSTW2008NLO~\cite{Sherstnev:2007nd} and NNPDF2.3~\cite{Ball:2012cx}
are used or the CT10 PDF set parameters are varied within their uncertainties. 
The largest variation in acceptance is taken as a systematic uncertainty. 
The uncertainty on the $H \rightarrow b\bar b$ branching ratio, 3.2\%~\cite{Djouadi:1997yw}, 
is also accounted for.

The uncertainty on higher order QCD corrections to the $Z \rightarrow b\bar b$ yield is estimated by varying 
the factorisation and renormalisation scales around the nominal value in the manner described above. 
It is found to be about 40-50\%, depending on the BDT category, out of which about 25\% is correlated. 
These correlated and uncorrelated uncertainties are used to constrain the $Z$ yield in the fit. 
This process results in about 20-25\% to the total uncertainty on the Higgs yield.

\section{Statistical procedure and results}\label{sec:StatisticalProcedure}

A statistical fitting procedure based on the RooStats framework~\cite{Moneta:2010pm,Verkerke:2003ir} 
is used to estimate the Higgs signal strength, $\mu$, from the data, where 
$\mu$ is the ratio of the measured signal yield to the SM prediction.
A binned likelihood function is constructed as the product of Poisson-probability terms of the bins  
in the $m_{bb}$ distributions, and of the four different BDT categories.

The impact of systematic uncertainties on the signal and background expectations, 
presented in Section~\ref{sec:Systematics}, is described 
by a vector of 
nuisance parameters (NPs), $\vec{\theta}$. The expected numbers of signal and background 
events in each bin and category are functions of $\vec\theta$.  For each NP with an a priori 
constraint, the prior is taken into account as a Gaussian constraint in the likelihood.
The NPs associated with uncertainties 
in the shape and normalisation of the non-resonant background events, which do not have priors, 
are determined from the data.

The test statistic $q_\mu$ is constructed according to the profile-likelihood ratio: 
\begin{equation}
q_\mu = 2\ln(\cal{L}(\mu, \vec\theta_\mu)/\cal{L}(\hat{\mu}, \vec{\hat{\theta}})), 
\end{equation}
where $\hat{\mu}$ and $\vec{\hat{\theta}}$ are the parameters that maximise the likelihood, 
and $\vec\theta_\mu$ are the nuisance parameter values that maximise the likelihood for a given $\mu$. 
This test statistic is used both to measure the compatibility of the background-only model with 
the data, and to determine exclusion intervals using the CL$_{S}$ method~\cite{Read:2002hq,Cowan:2010js}.

The robustness of the fit is validated by generating pseudo-data and estimating the number of
signal events for various values of $\mu$.
The results of the fit in the four categories are shown in Figure~\ref{fig:catfit_pl}. 
The $Z$ yield is constrained to the SM prediction within its theoretical uncertainty,
using four independent constraints in the four BDT regions (uncorrelated terms) 
and a common constraint (correlated term)
as described in Section~\ref{sec:SystematicsTheory}. 
The ratios of $Z$ yields to the SM predictions ($\mu_Z$) are found to be compatible in all of the four BDT regions.
Combined over the four categories, the fit further constrains $\mu_Z$ to $0.7 \pm 0.2$.

The combined Higgs signal strength is $-0.8 \pm2.3$, where the uncertainty 
includes both the statistical ($\pm1.3$) and systematic (+1.8/$-$1.9) components. 
The breakdown of the systematic uncertainty on the estimated signal strength is given in 
Table~\ref{tab:muhat}. The correlation coefficient between the combined $\mu$ and the combined $\mu_Z$ 
is found to be 0.22. 
In the absence of a signal, the limit on the Higgs signal strength at 95\% confidence level (CL) is expected to 
be 5.4. When Standard Model production is assumed, the expected limit is found to be 5.7.
The observed limit is $4.4$.

The compatibility between the measured $Z$ yield and its SM prediction is alternatively tested 
by removing its a priori constraint from the fit. In this case a value of $\mu_{Z} = 0.3 \pm 0.3$ 
is extracted from the fit, to be compared to the theory prediction of $1.0 \pm 0.4$. The absence of 
the $Z$ constraint modifies the combined Higgs signal strength slightly, to $-0.5 \pm 2.3$.

\begin{figure}[ht!]
 \centering
 \subfloat[]{\includegraphics[width=0.5\linewidth]{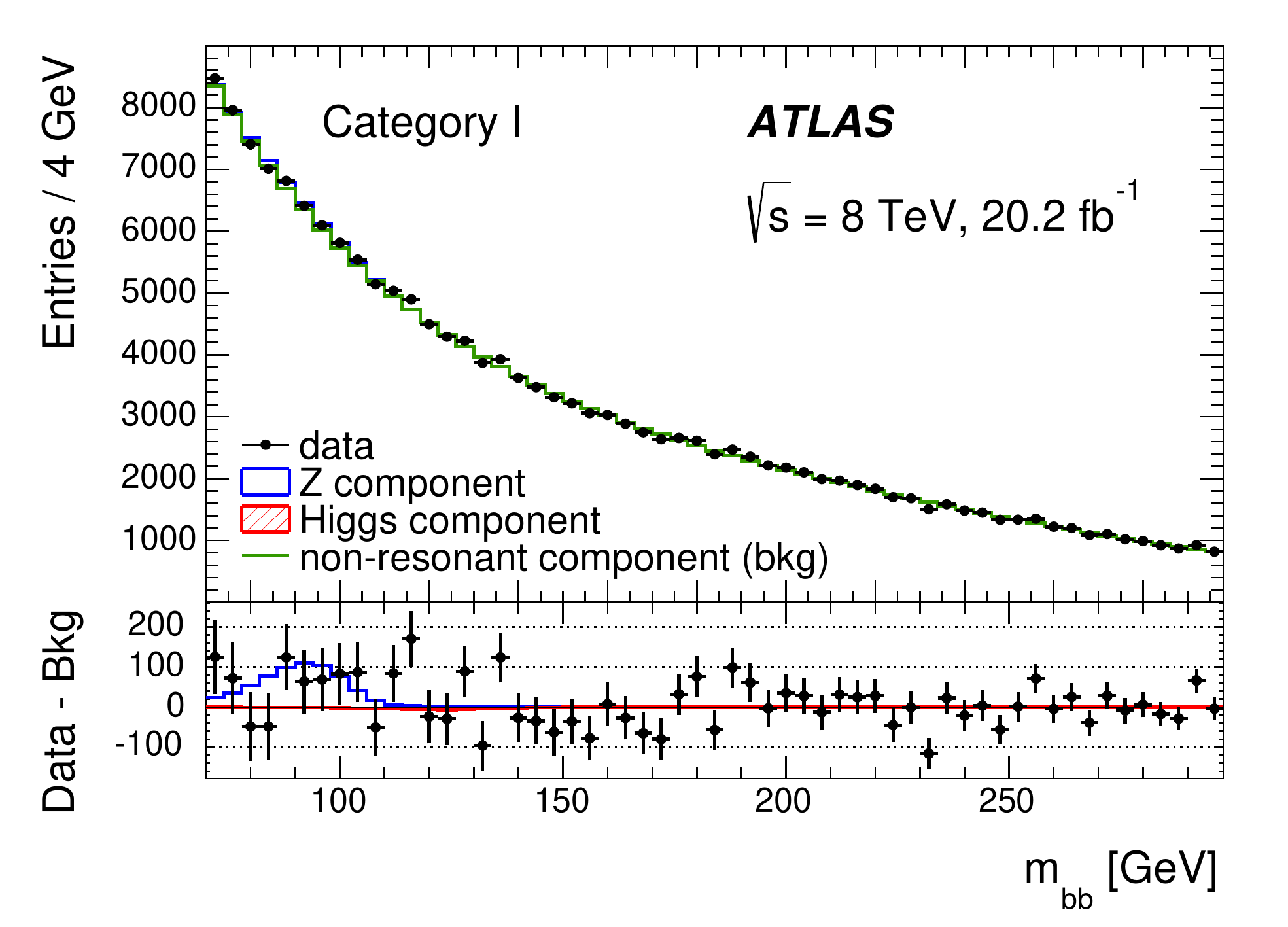}}
 \subfloat[]{\includegraphics[width=0.5\linewidth]{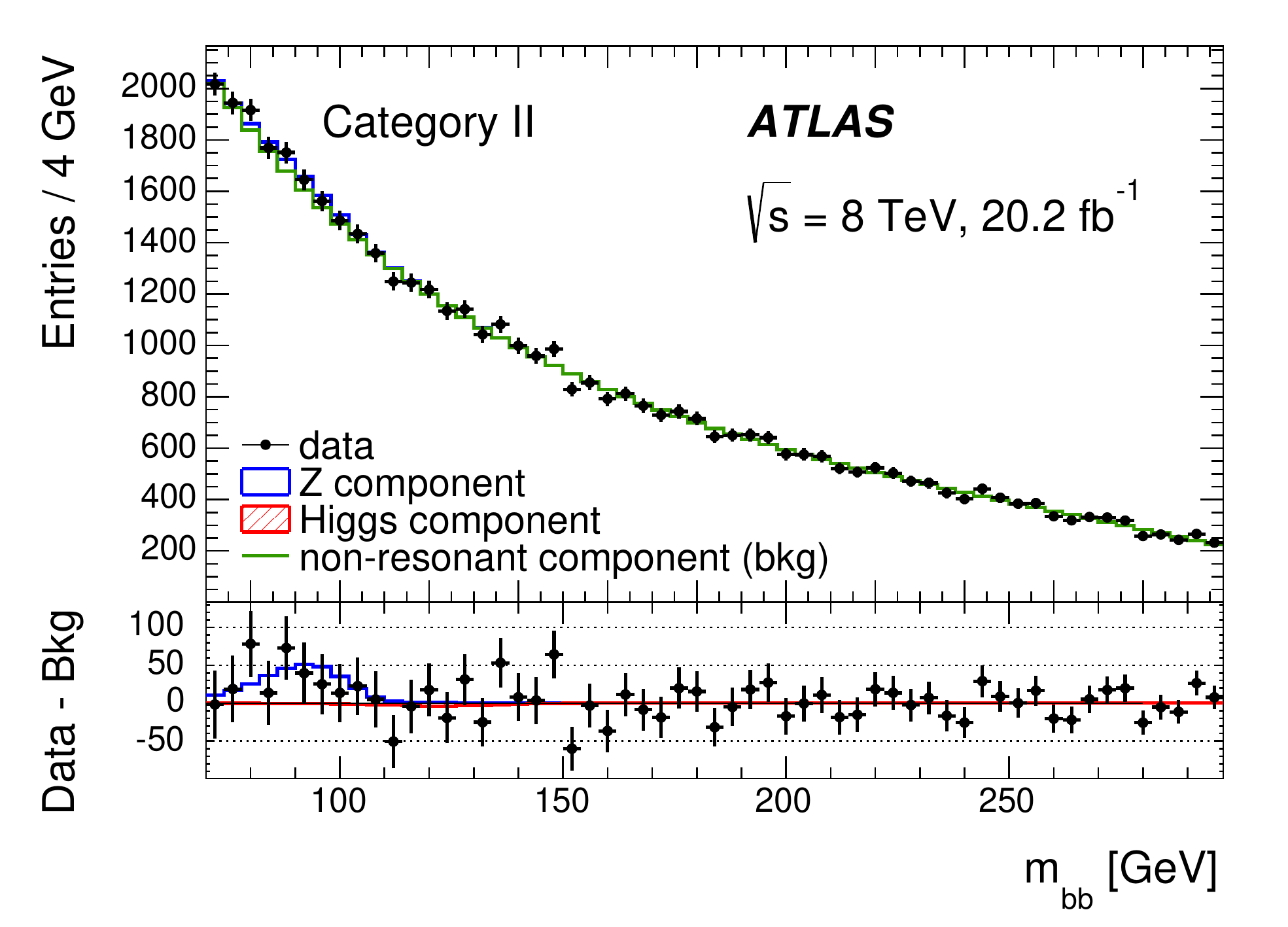}} \\
 \subfloat[]{\includegraphics[width=0.5\linewidth]{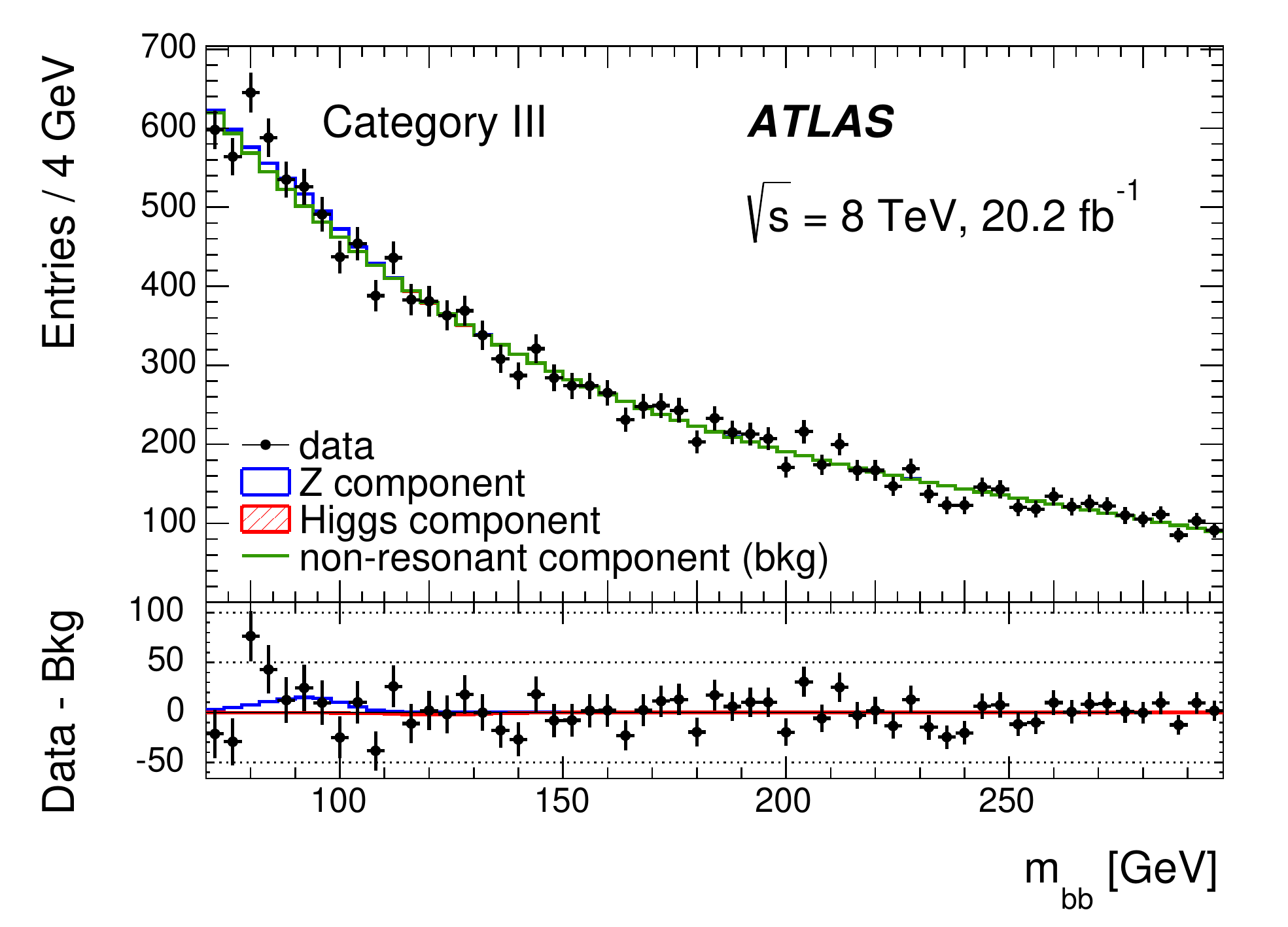}}
 \subfloat[]{\includegraphics[width=0.5\linewidth]{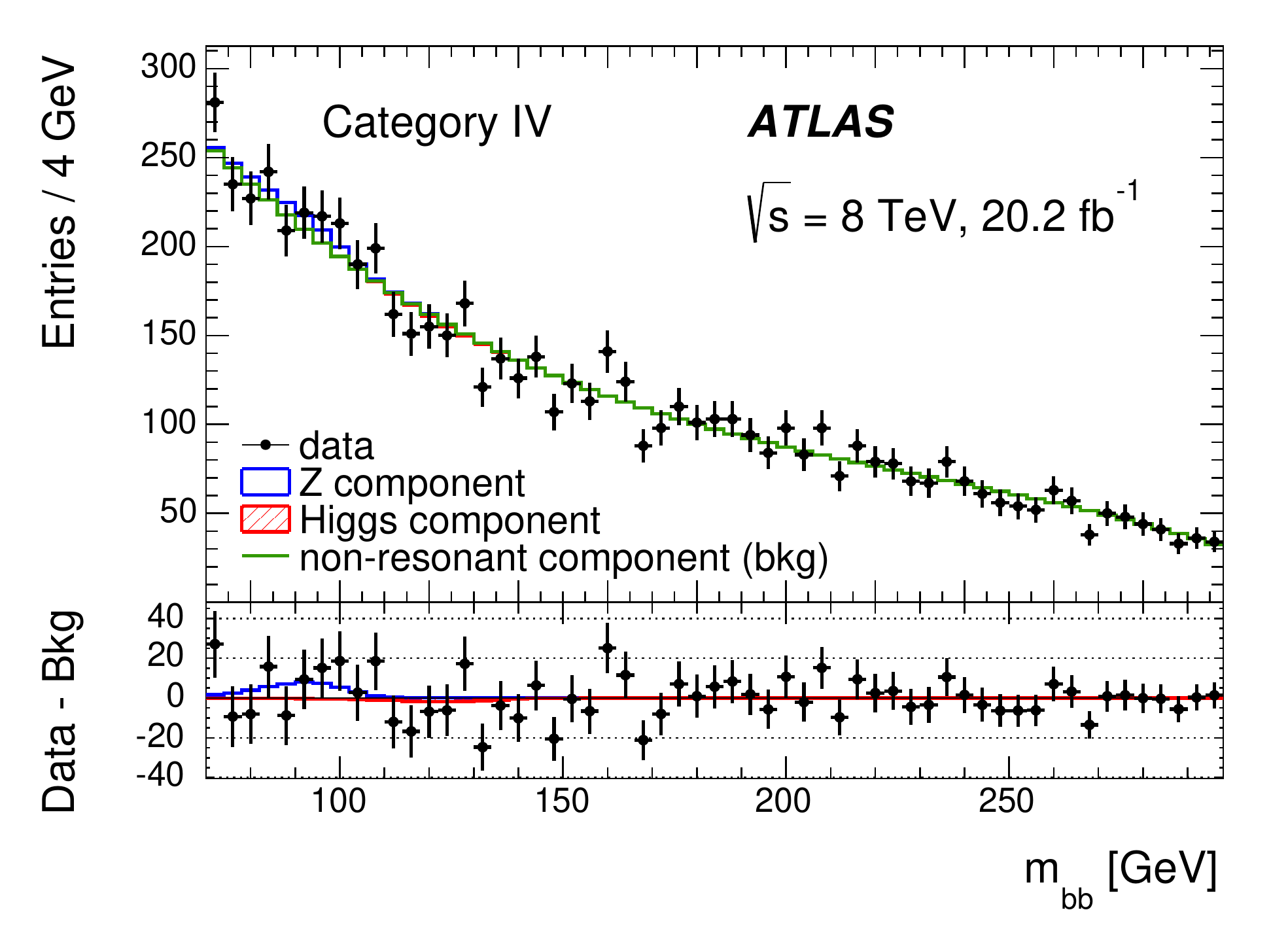}} \\
 \caption{
   Results of the profile-likelihood fit to the $m_{bb}$ distributions in the four BDT categories.
   The points represent the data, and the histograms represent the non-resonant 
   background, $Z$, and Higgs contributions.
   In the lower panels, the data after subtraction of the non-resonant background (points) are 
   compared with the fit to the $Z$ (open histogram) and Higgs (shaded histogram) contributions.
} 
\label{fig:catfit_pl}
\end{figure}

\begin{table}[!h]
\begin{center}
\caption{
  Summary of uncertainties on the Higgs signal strength for the MVA analysis, and 
  for the cut-based analysis.
  They are estimated 
  at the central values of the signal strength, $\mu$ = $-$0.8 and $-$5.2 
  for the MVA and cut-based analyses, respectively. 
  The two systematic uncertainties accounting for non-resonant
  background modelling are strongly correlated. Their combined
  value for the MVA analysis is 1.8.
}
\label{tab:muhat}
\begin{tabular}{|l|l|c|c|}
\hline 
\multicolumn{2}{|l|}{Source of uncertainty}
& \multicolumn{2}{c|}{Uncertainty on $\mu$} \\ \cline{3-4}
        \multicolumn{2}{|c|}{} & MVA & Cut-based \\ \hline
\hline 
\multirow{2}{*}{Experimental uncertainties} &  Detector-related     &  +0.2/$-$0.3 & +1.6/$-$1.2 \\ \cline{2-4} 
                         &  MC statistics        &  $\pm0.4$ & $\pm 0.1$ \\ \hline
\multirow{2}{*}{Theoretical uncertainties}  &  MC signal modelling  &  $\pm0.1$ & $\pm 1.3$ \\ \cline{2-4}
                         &  $Z$ yield & +0.6/$-$0.5 & $\pm 1.4$ \\ \hline
\multirow{2}{*}{Non-resonant background modelling}     &  Choice of function &  $\pm 1.0$ & $\pm 1.0$ \\ \cline{2-4}
                                         &  Sideband statistics & $\pm1.7$ & \multirow{2}{*}{$\pm 3.7$}\\ \cline{1-3}
\multicolumn{2}{|l|}{Statistical uncertainties}   &  $\pm 1.3$ & 
\\ \hline 
\hline
\multicolumn{2}{|l|}{Total} & $\pm 2.3$ & +4.6/$-$4.4 \\ \hline
\end{tabular}
\end{center}
\end{table}

\section{Cut-based analysis}\label{sec:CutBased}

An alternative analysis is performed based on kinematic cuts.
While the MVA performs 
a simultaneous fit to the $m_{bb}$ distributions of the four samples categorised 
by the BDT response, the cut-based analysis performs a fit to one $m_{bb}$ 
distribution of the entire sample in the mass range between $70$~$\gev$ and $300$~$\gev$. 
Events are required to satisfy kinematic criteria featuring the VBF Higgs final state. 
Events must not have any additional jet with
$\pT > 25$~$\gev$ and $|\eta| < 2.4$, and must satisfy $|\Delta \eta_{JJ}|>3.0$ and 
$m_{JJ} > 650$~$\gev$. 
Figure~\ref{fig:mbb_cutbased} shows the $m_{bb}$ distribution of 32906 events in the data 
that satisfy the selection criteria.
The number of signal events in the data is 
expected to be 68.8, with about 15\% coming from ggF production. This can be 
compared to 158.9 events in the MVA, as obtained by summing the corresponding 
numbers in Table~\ref{tab:cat} over the four categories, where about 28\% comes from ggF production.

\begin{figure}[!h]
 \centering
 \includegraphics[width=0.5\textwidth]{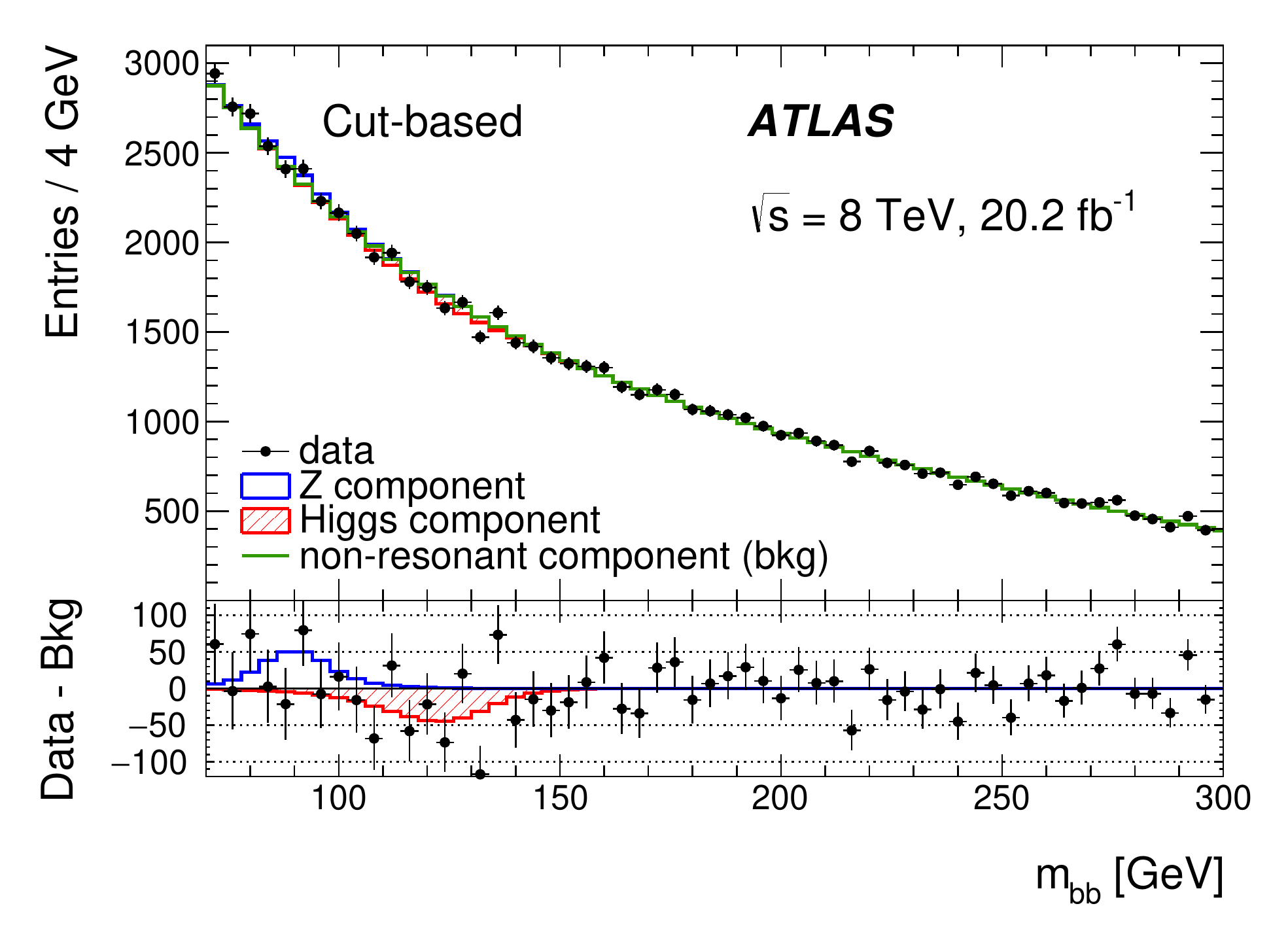}
 \caption{
Distribution of $m_{bb}$ for events selected in the cut-based analysis.
The points represent the data, 
and the histograms represent the non-resonant background, $Z$, and Higgs
contributions. In the lower panel, the data after subtraction of the
non-resonant background (points) are compared with the fit to the $Z$ (open histogram) and Higgs
(shaded histogram) contributions. The Higgs yield extracted from the fit is consistent with zero.
 }
 \label{fig:mbb_cutbased}
\end{figure}

The cut-based analysis uses an unbinned maximum likelihood fit.
The resonance shapes of the $m_{bb}$ distributions for the Higgs and $Z$ 
events are determined by a fit to a Bukin function~\cite{Bukin} using MC events. 
The analytic functions describing the non-resonant background are studied by using 
events that satisfy the pre-selection criteria described in Section~\ref{sec:EventSelection}. 
A fourth-order polynomial is chosen as the nominal function and a fifth-order 
polynomial is chosen as the alternative function.
 
The Higgs yield is left free to vary, but the $Z$ yield is fixed to its SM prediction. 
The robustness of the fit is validated by generating pseudo-data and 
constructing pulls of the estimated number of Higgs events for 
various values of $\mu$. The fit results are presented in Figure~\ref{fig:mbb_cutbased}.
The Higgs signal strength is measured to be 
$\mu = -5.2 \pm 3.7{\rm (stat.)}^{+2.7}_{-2.5}{\rm (syst.)}$, where the statistical 
uncertainty includes the statistical uncertainty on the non-resonant background 
modelling (see Table~\ref{tab:muhat}). The sources of systematic uncertainty are 
the same as those for the MVA analysis as described in Section~\ref{sec:Systematics}
and are summarised in Table~\ref{tab:muhat}. 
The uncertainties on $\mu$ are estimated as the changes in $\mu$ when the sources are varied within 
their uncertainties. Higher-order corrections to the $Z$ samples and to the signal samples, 
the choice of function describing the non-resonant 
background, and the jet energy scale are the dominant sources of systematic 
uncertainty, each contributing about 40--50\% to the total systematic 
uncertainty on the Higgs signal strength.  
The magnitudes of experimental and theoretical uncertianties are scaled with the central value of $\mu$, 
as illustrated in Table~\ref{tab:muhat} except for the case of the MC statistical uncertainty.
This is due to the fact that the MVA divides the MC samples into four categories, and 
uses the signal $m_{bb}$ distribution directly in the fit as a template 
while the cut-based analysis uses an interpolated function.
The upper limit on the strength is found 
to be 5.4 at the 95\% CL, which can be compared to the expected limit values
of 8.5 in the background-only hypothesis and 9.5 if Standard Model production is assumed. 
These results are consistent with those of the MVA. As expected, the cut-based analysis 
is less sensitive than the MVA.

\section{Summary}\label{sec:Summary}

A search for the Standard Model Higgs boson produced by vector-boson fusion 
and decaying into a pair of bottom quarks is presented. The dataset 
analysed corresponds to an integrated luminosity of 20.2~$\ifb$ from $pp$ 
collisions at $\sqrt{s}= 8$~$\tev$, recorded by the ATLAS experiment during 
Run 1 of the LHC. Events are selected using the distinct final state of the 
VBF $H \rightarrow b\bar b$ signal, which is the presence of four energetic jets: two 
$b$-jets from the Higgs boson decay in the central region of the 
detector and two jets in the forward/backward region. 
To improve the sensitivity, a multivariate analysis is used, 
exploiting the topology of the VBF Higgs final state and the properties
of jets. The signal yield is estimated by performing a fit 
to the invariant mass distribution of the two $b$-jets in the range 
$70 < m_{bb} < 300$~$\gev$ and assuming a Higgs boson mass of $125$~$\gev$. 
The ratio of the Higgs signal yield to the SM prediction is measured to be 
$\mu = -0.8 \pm 1.3{\rm (stat.)}^{+1.8}_{-1.9}{\rm (syst.)} = -0.8 \pm 2.3$.
The upper limit on $\mu$ is observed to be $\mu$ = $4.4$ at the 95\% CL, 
which should be compared to the expected limits of 5.4
in the background-only hypothesis and 5.7
if Standard Model production is assumed.
An alternative analysis is performed using kinematic selection criteria  
and provides consistent results: $\mu = -5.2 ^{+4.6}_{-4.4}$
and a 95\% CL upper limit of 5.4.

\section*{Acknowledgements}

We thank CERN for the very successful operation of the LHC, as well as the
support staff from our institutions without whom ATLAS could not be
operated efficiently.

We acknowledge the support of ANPCyT, Argentina; YerPhI, Armenia; ARC, Australia; BMWFW and FWF, Austria; ANAS, Azerbaijan; SSTC, Belarus; CNPq and FAPESP, Brazil; NSERC, NRC and CFI, Canada; CERN; CONICYT, Chile; CAS, MOST and NSFC, China; COLCIENCIAS, Colombia; MSMT CR, MPO CR and VSC CR, Czech Republic; DNRF and DNSRC, Denmark; IN2P3-CNRS, CEA-DSM/IRFU, France; GNSF, Georgia; BMBF, HGF, and MPG, Germany; GSRT, Greece; RGC, Hong Kong SAR, China; ISF, I-CORE and Benoziyo Center, Israel; INFN, Italy; MEXT and JSPS, Japan; CNRST, Morocco; FOM and NWO, Netherlands; RCN, Norway; MNiSW and NCN, Poland; FCT, Portugal; MNE/IFA, Romania; MES of Russia and NRC KI, Russian Federation; JINR; MESTD, Serbia; MSSR, Slovakia; ARRS and MIZ\v{S}, Slovenia; DST/NRF, South Africa; MINECO, Spain; SRC and Wallenberg Foundation, Sweden; SERI, SNSF and Cantons of Bern and Geneva, Switzerland; MOST, Taiwan; TAEK, Turkey; STFC, United Kingdom; DOE and NSF, United States of America. In addition, individual groups and members have received support from BCKDF, the Canada Council, CANARIE, CRC, Compute Canada, FQRNT, and the Ontario Innovation Trust, Canada; EPLANET, ERC, FP7, Horizon 2020 and Marie Sk{\l}odowska-Curie Actions, European Union; Investissements d'Avenir Labex and Idex, ANR, R{\'e}gion Auvergne and Fondation Partager le Savoir, France; DFG and AvH Foundation, Germany; Herakleitos, Thales and Aristeia programmes co-financed by EU-ESF and the Greek NSRF; BSF, GIF and Minerva, Israel; BRF, Norway; Generalitat de Catalunya, Generalitat Valenciana, Spain; the Royal Society and Leverhulme Trust, United Kingdom.

The crucial computing support from all WLCG partners is acknowledged gratefully, in particular from CERN, the ATLAS Tier-1 facilities at TRIUMF (Canada), NDGF (Denmark, Norway, Sweden), CC-IN2P3 (France), KIT/GridKA (Germany), INFN-CNAF (Italy), NL-T1 (Netherlands), PIC (Spain), ASGC (Taiwan), RAL (UK) and BNL (USA), the Tier-2 facilities worldwide and large non-WLCG resource providers. Major contributors of computing resources are listed in Ref.~\cite{ATL-GEN-PUB-2016-002}.

\clearpage

\printbibliography
\clearpage

 \newpage 
\begin{flushleft}
{\Large The ATLAS Collaboration}

\bigskip

M.~Aaboud$^{\rm 136d}$,
G.~Aad$^{\rm 87}$,
B.~Abbott$^{\rm 114}$,
J.~Abdallah$^{\rm 65}$,
O.~Abdinov$^{\rm 12}$,
B.~Abeloos$^{\rm 118}$,
R.~Aben$^{\rm 108}$,
O.S.~AbouZeid$^{\rm 138}$,
N.L.~Abraham$^{\rm 150}$,
H.~Abramowicz$^{\rm 154}$
% ...
\bigskip
\\
% ...
$^{12}$ Institute of Physics, Azerbaijan Academy of Sciences, Baku, Azerbaijan\\
% ...
$^{65}$ University of Iowa, Iowa City IA, United States of America\\
% ...
$^{87}$ CPPM, Aix-Marseille Universit{\'e} and CNRS/IN2P3, Marseille, France\\
% ...
$^{108}$ Nikhef National Institute for Subatomic Physics and University of Amsterdam, Amsterdam, Netherlands\\
% ...
$^{114}$ Homer L. Dodge Department of Physics and Astronomy, University of Oklahoma, Norman OK, United States of America\\
% ...
$^{118}$ LAL, Univ. Paris-Sud, CNRS/IN2P3, Universit{\'e} Paris-Saclay, Orsay, France\\
% ...
$^{136}$ $^{(a)}$ Facult{\'e} des Sciences Ain Chock, R{\'e}seau Universitaire de Physique des Hautes Energies - Universit{\'e} Hassan II, Casablanca; $^{(b)}$ Centre National de l'Energie des Sciences Techniques Nucleaires, Rabat; $^{(c)}$ Facult{\'e} des Sciences Semlalia, Universit{\'e} Cadi Ayyad, LPHEA-Marrakech; $^{(d)}$ Facult{\'e} des Sciences, Universit{\'e} Mohamed Premier and LPTPM, Oujda; $^{(e)}$ Facult{\'e} des sciences, Universit{\'e} Mohammed V, Rabat, Morocco\\
% ...
$^{138}$ Santa Cruz Institute for Particle Physics, University of California Santa Cruz, Santa Cruz CA, United States of America\\
% ...
$^{150}$ Department of Physics and Astronomy, University of Sussex, Brighton, United Kingdom\\
% ...
$^{154}$ Raymond and Beverly Sackler School of Physics and Astronomy, Tel Aviv University, Tel Aviv, Israel\\
% ...
$^{a}$ Also at Department of Physics, King's College London, London, United Kingdom\\
$^{b}$ Also at Institute of Physics, Azerbaijan Academy of Sciences, Baku, Azerbaijan\\
$^{c}$ Also at Novosibirsk State University, Novosibirsk, Russia\\
$^{d}$ Also at TRIUMF, Vancouver BC, Canada\\
$^{e}$ Also at Department of Physics {\&} Astronomy, University of Louisville, Louisville, KY, United States of America\\
% ...
\end{flushleft}

\end{document}